\documentclass[utf8,aps,twocolumn,pre]{revtex4-2}

\usepackage{graphicx}
\usepackage{epstopdf}
\usepackage{listings}	
\usepackage{amssymb}
\usepackage{amsmath}
\usepackage{amsthm}
\usepackage[extension=xxx]{hyperref}
\usepackage{color}
\usepackage[utf8]{inputenc}
\usepackage{lipsum}

\def\be{\begin{equation}}
\def\ee{\end{equation}}
\def\bea{\begin{eqnarray}}
\def\eea{\end{eqnarray}}
\newcommand*\diff{\mathop{}\!\mathrm{d}}

\newcommand{\del}{\partial}

\newcommand{\open}{\left(}
\newcommand{\close}{\right)}

\begin{document}

\author{Arthur Alexandre}
\affiliation{Institute of Bioengineering, School of Life Sciences, École Polytechnique Fédérale de Lausanne (EPFL), CH-1015 Lausanne, Switzerland}
\affiliation{SIB Swiss Institute of Bioinformatics, CH-1015 Lausanne, Switzerland}
\author{Simone Cicolini}
\affiliation{Department of Genetics and Evolution, University of Geneva, Quai Ernest-Ansermet 30, 1205 Geneva, Switzerland.}
\author{Nicola Dietler}
\affiliation{Institute of Bioengineering, School of Life Sciences, École Polytechnique Fédérale de Lausanne (EPFL), CH-1015 Lausanne, Switzerland}
\affiliation{SIB Swiss Institute of Bioinformatics, CH-1015 Lausanne, Switzerland}
\author{Andrew Callan-Jones}
\affiliation{Laboratoire Matière et Systèmes Complexes, Université de Paris Cité, Paris, France.}
\author{Dennis Wörthmüller}
\affiliation{Institut Curie, Université PSL, Sorbonne Université, CNRS UMR168, Physique des Cellules et Cancer, Paris, France}
\author{Serge Dmitrieff}
\email{serge.dmitrieff@ijm.fr}
\affiliation{Université Paris Cité, CNRS, Institut Jacques Monod, F-75013 Paris, France}

\title{Microscopic origin of macroscopic contractility in actin-myosin active gel models.}
\begin{abstract}
Actin filaments, crosslinkers and myosin molecular motors form contractile networks. For instance, the cell cortex is a thin network below the cell membrane ; contraction of the cell cortex allows cells to round up during cell division. Contractile actin-myosin networks are often represented at large scale by continuous theories such as active gel models. However, experimental perturbations are microscopic while parameters in continuous models are macroscopic, thus making those models hard to falsify experimentally.
Here we use numerical simulations, in which we can access both microscopic and macroscopic quantities, to show that active gel models can indeed be applied to describe contractile actin. We predict that contractile stress should scale linearly with actin  density, which is confirmed by numerical simulations. Moreover, we can accurately predict how the contractile stress depends on motor properties such as unloaded speed and stall force.
\end{abstract}

\maketitle

Actin networks are
 an element of the cell cytoskeleton, a dynamic assembly of filaments and associated proteins, that have a wide range of cellular functions. In particular, actin networks play a role in muscle contraction, intra-cellular transport, and cellular mechanics \cite{letort2015dynamic}. 

Actin filaments are polar, dynamic helicoidal filaments. Their persistence length is of the order of $10\mu m$ (also the typical size of a cell), and their length range from a tens of nanometers to several micrometers \cite{letort2015dynamic}. Motor proteins associated with actin can consume energy by hydrolyzing Adenosine TriPhosphate (ATP) to move directionally along actin filaments, producing forces of the order of several picoNewton  \cite{kolomeisky2007molecular}. This hydrolysis releases tens of $k_B T$, allowing actin filaments to be deformed on scales much smaller than their persistence length.

Actin-myosin networks are known to usually be contractile, even when lacking the structured organization
 of muscle sarcomeres. A canonical example is that of the cell cortex, a thin layer of actin below the cell membrane. The contraction of the cortex is responsible for cell rounding, a key event in the context of cell division \cite{salbreux2012actin}.

Directional motor movement on filaments could yield extensile and contractile configurations. However, while actin filaments bear load and hardly stretch under tension, they buckle under compression if the load is larger than the typical Euler buckling load \cite{lenz2012contractile}. In the presence of crosslinkers, this non-linear elasticity is known to break the symmetry towards contractility, although other mechanisms such as zippering could also be involved \cite{lenz2014geometrical}. 

The most popular formalism for a macroscopic description of actin-myosin networks is the "active gel" formalism, which represents them as a continuous viscoelastic medium, in which activity is introduced as an additional term $\sigma_a$ in the constitutive equation for the stress \cite{juelicher2007active}. Active gel formalism has been extensively used to model contractile actin networks, with different choices for the terms kept in the visco-elastic stress \cite{joanny2007hydrodynamic}. The macroscopic quantities (such as the viscoelastic moduli and timescales, or the contractility) depend in complex ways on the microscopic details (such as actin and myosin density, ATP availability, etc.). Therefore, there is little to no quantitative comparison between active gel models and experiments to ensure the terms kept in the theory are relevant. Here, we intend to bypass this limitation by measuring these macroscopic properties, as well as the stress, in detailed numerical simulations that explicitly represent the microscopic components : actin filaments, crosslinkers, and motors.



\paragraph*{Time evolution of contractile actin networks}

Actin gels are usually assumed to follow the Maxwell equation, with $\sigma^v$ the visco-elastic stress tensor :
\be 
\frac{\partial}{\partial t} \sigma_{ij}^v = E v_{ij} - \frac{1}{\tau}\sigma_{ij}^v
\ee
In which $v_{ij} = \partial_i v_j +\partial_j v_i$ is the symmetric strain rate tensor \cite{juelicher2007active}. In the quasi-stationary assumption $\partial_t \sigma^v \approx 0$ yielding  $\sigma_{ji}^v = \eta v_{ij} $, with $\eta = E \tau$.

The active contractile stress $\sigma^a$ is often assumed to be a negative pressure, i.e. $\sigma^a \propto \delta_{ij}$. We will assume the system to be infinitely compressible, i.e. there is no additional pressure term from network compression. Moreover, we expect the network viscosity to be much larger than that of the surrounding fluid, allowing us to neglect filament friction. In the absence of other external forces, force balance yields : 
\be 
\nabla .  \left( \sigma^v  +  \sigma^a \right) = 0
\ee
Solving this equation should yield a solution for $v_{ij}$ and thus for the velocity field $v$. Then we can use mass conservation to obtain the time evolution of the system :
\be 
\partial_t \phi = - \nabla . \left( \mathbf{v} \phi \right)
\ee

Here we will focus on systems in two spatial dimensions (2D), assuming radial symmetry. Calling $v$ the radial velocity field for simplicity, we can thus write force balance as a function of radius $r$ :
\begin{eqnarray}
2 \partial_r (\eta \partial_r v) + \frac{2\eta}{r} \left( \partial_r v - \frac{v}{r}  \right) = - \partial_r 
\sigma_{rr}^a  \label{eq_stress}\\
\partial_t  \phi = - \frac{1}{r} \partial_r \left( r v \phi \right) \label{eq_mass}
\end{eqnarray}
These equations will not easily yield an analytical solution because $\sigma^a$ and $\eta$ are expected to  depend on actin density. Rather, we will use a simple ansatz to find a possible solution.

First, we considered a constant-density region of actin with radial symmetry. In this case, equations \ref{eq_stress}--\ref{eq_mass} yield spatially constant contraction rate $\alpha$ : 
\begin{eqnarray}
 v(r,t) = - \alpha(t) r \label{alpha_v}\\
\partial_t \phi(t) = 2 \alpha(t) \phi(t)  \label{alpha_phi}
\end{eqnarray}

\begin{figure}[t]
	\includegraphics[width=9cm]{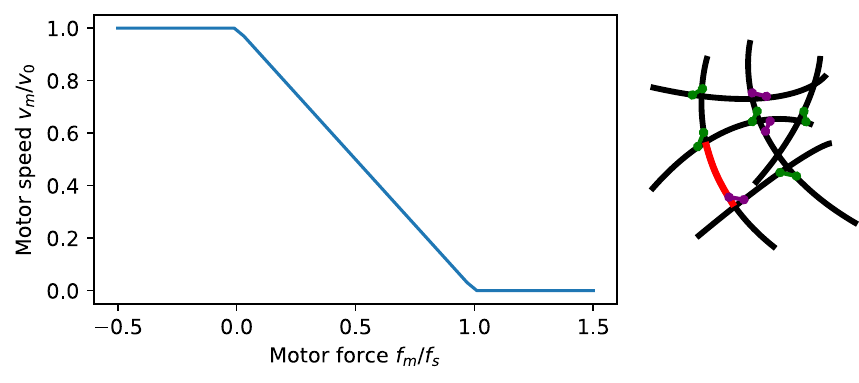} 	
	\caption{\label{illus_network} \small Left : motor velocity as a function of force (where $f_m > 0$ opposes movement). Right : schematic illustration of a small fraction of the network. Actin filaments are in black, motors, in purple, crosslinkers in green. A basic force-producing unit (a span of filament between a motor and a crosslinker) is highlighted in red.
	}
\end{figure}

In the absence of external forces on the system, we assume $\sigma_{rr}$ to be zero at the domain boundary $r=R_b(t)$.
	Assuming Eq. \ref{alpha_v} to hold and the actin patch interface to be small, the boundary condition on $\sigma_{rr}$ yields :
\be
 \alpha(t) = \frac{\sigma^a_{rr}(t)}{2 \eta (t) }  \label{pred_alpha}
\ee
Generically, $\sigma^a_{rr}$ and $\eta$ should depend on the density $\phi$, and so should the contraction rate $\alpha$. 

If actin networks indeed behave as active gels, they should follow this predicted behaviour. To check this, we performed numerical simulations of contractile actin using the open-source platform {\em Cytosim} \cite{nedelec2007collective,lugo2023typical}. In this platform, filaments are discretized into chains of vertices, for which the Langevin equation is solved by an implicit scheme. 
Vertex displacement is projected to impose a constant segment length between vertices to make filaments incompressible. We also impose filaments to live in a two-dimensional plane. We discard steric repulsions between filaments because they yield unphysical behaviour in 2D compared to a very thin network ; this is consistent with our assumption of infinite compressibility. While filament drag is implemented directly in the Langevin equations, hydrodynamic interactions between filaments are not included in the simulation.  Here we follow a simulation configuration very similar to previous studies of actin contraction \cite{belmonte2017theory}. 

 Filaments are flexible, with a bending rigidity (here $\kappa = 0.075\, pN \mu m^2$) penalizing the angle between adjacent segments. Motors and crosslinkers are represented as elastic springs (here of stiffness $250 \, pN/\mu m$ and resting length $10 \, nm$) that are attached anywhere along a filament arclength. Both crosslinkers and motors can bind to filaments with a rate $k_\text{on}=10\, s^{-1}$, within a $10\, nm$ distance of a filament, and unbind at a rate $k_\text{off}=0.1\, s^{-1}$. Unbinding is implemented using the Gillespie algorithm, while binding results in a probability $k_\text{on} dt$ to bind for crosslinkers and motors within a range $a$ of a filament at each time step $dt$. 

While crosslinkers have a fixed position, motors move on filaments with a speed $v_m$, towards the plus end. Thermodynamics imposes $v_m$ to decrease with the force $f_m$ applied on the motor in the direction of the movement (with the sign convention that $f_m >0$ \emph{opposes} motor movement) \cite{keller2000mechanochemistry}. Here we approximate the force-velocity relationship of motors with : \cite{belmonte2017theory}
\begin{eqnarray}
v_m = v_0 \left(1 - \frac{f_m}{f_{s}} \right) \qquad \text{if} \quad 0 \le f_m \le f_s \label{motor_1} \\
v_m = 0  \qquad \text{ if}  \qquad f_m \ge f_s  \label{motor_2}\\
v_m = v_0 \qquad \text{if} \qquad f_m \le \, 0 \, \label{motor_3}
\end{eqnarray}
With $v_0$ and $f_s$ the motor unloaded speed and stall force, Fig. \ref{illus_network}, top. Here, we chose to have motors detach from a filament when they reach the filament end.

\begin{figure}[t]
	\includegraphics[width=6cm]{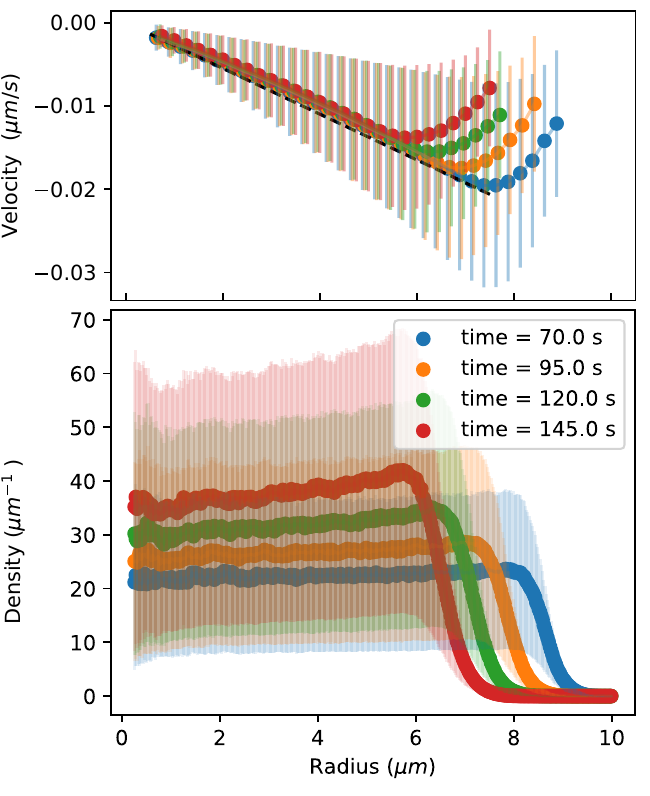} 	
	\includegraphics[width=6cm]{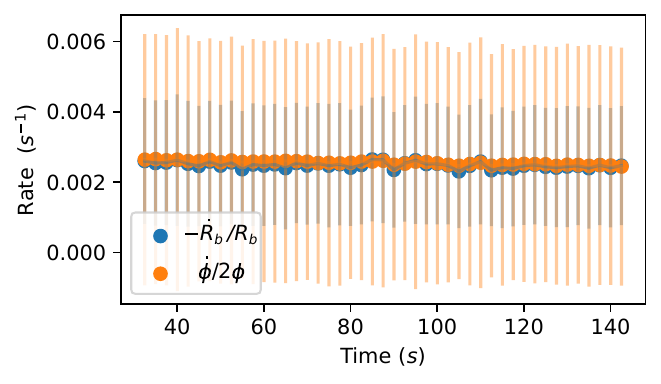} 
	\caption{\label{illus_active_gel} \small Top : velocity $v$ as a function of radius $r$ at several times ; dashed line is a linear guideline. Middle : density (defined as actin length per unit surface, i.e. in $\mu m / \mu m^2$) as a function of radius $r$ at several times. Bottom : contraction rate measured according to Eq. \ref{alpha_v} (yellow) and Eq. \ref{alpha_phi} (blue). Here motors have $f_s = 6.7 \, pN$ and $v_0=0.07 \mu m / s$. All results are averages from 72 simulations ; error bars are standard deviation.}
\end{figure}

We simulated networks of $5000$ filaments of length $1.5\mu m$, $10000$ motors and $40000$ crosslinkers, initially distributed homogeneously on a disc of radius $R=10 \, \mu m$. We find that the initially homogeneous network of actin, motors, and crosslinkers contracts regularly with time, with a near-constant plateau density in the center and a near-zero density outside, Fig. \ref{illus_active_gel}, middle. Because of mass conservation, the plateau density in the network increases as the network radius decreases. As predicted by our simple ansatz, the velocity $v$ scales linearly with $r$, Fig. \ref{illus_active_gel}, top. The contraction rate measured by monitoring network radius matches the one obtained by measuring $\dot{\phi}/2\phi$, Fig. \ref{illus_active_gel}, bottom. Importantly, the contraction rate is constant in time, indicating that the viscosity and contractility have the same dependence on density, Eq. \ref{pred_alpha}.


\paragraph*{Emergence of macroscopic contractility}
We then aimed at predicting how the contractile stress depends on density. Previous work proposed a theory predicting the contractile rate (up to a prefactor) of networks according to the density of actin filaments, motors, and crosslinkers \cite{belmonte2017theory}. For this, they assumed the force-producing unit to be the span of a filament between two intersections. Such a unit will produce force if there is (at least) one crosslinker at one intersection, and (at least) a motor, but no crosslinker, at the other intersection, Fig. \ref{illus_network}, right. Calling $a$ the length of the unit (i.e. the typical distance between two intersections), they showed that the contraction rate should scale as as :
\begin{eqnarray}
\alpha \propto \frac{v_m}{a} \mathcal{C} \\
\mathcal{C} = P(\text{contractile}) - P(\text{extensile})
\end{eqnarray}
In which $\mathcal{C}$ is the effective contractility, and $P(\text{extensile}||\text{contractile})$ is the probability of the unit to be effectively extensile or contractile. For flexible filaments, most extensile configurations are spoiled by buckling, and $\mathcal{C}$ is dominated by contractility. Using simple geometrical arguments, $\mathcal{C}$ is predicted from the density of network elements \cite{belmonte2017theory}. Up to a prefactor, this theory predicts very accurately $\alpha$ as a function of the number of actin filaments, crosslinkers, and motors, but does not inform us on the dependence of $\sigma_{rr}^a$ on $\phi$.

For this, we used the finding that the active stress should be proportional to the density $\rho_u$ of active elements, times their force dipole moment \cite{ronceray2015connecting,ronceray2016fiber}. Here, the active unit is the portion of filaments between two intersections \cite{belmonte2017theory}, while the force dipole moment is the filament tension $t_u$ times the  distance $l_u$ between intersection, yielding $<\sigma^a> \approx \rho_u <t_u l_u >$. Assuming the network to be disordered and homogeneous, we thus obtain :
\be  
<\sigma^a> \approx \phi f_m \mathcal{C} \label{stress_fm}
\ee 
 With $\phi = \rho_u l_u$ the density of filaments, $f_m$ the motor force, and $\mathcal{C}$ the effective contractility. We then decided to test numerically this prediction that the contractile stress should be proportional to the linear actin density.
 
\begin{figure}[t]
	\includegraphics[width=7cm]{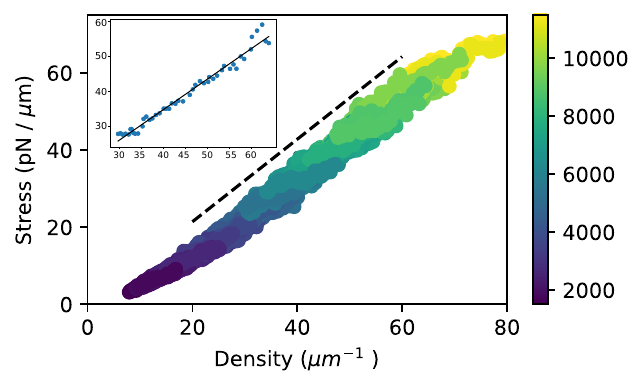} 	
	\caption{\label{illus_density} \small Measured stress $\sigma_{rr}^a$ as a function of density (defined as actin length per unit surface, i.e. in $\mu m / \mu m^2$) for multiple simulations, at different times. Each color corresponds to the initial number of filaments in the simulation. Dashed line : linear visual guide. Inset : measured stress as a function of actin density $\phi$ for one simulation with 5000 filaments ; one point corresponds to one timepoint ; black line : linear fit. }
\end{figure}

 In simulations, it is possible to measure the active stress by summing the tensions of the  filaments intersecting a line (in 2D), and projecting in the radial direction :
 \be  
 \sigma_{rr}^a (r) = \frac{1}{2 \pi r} \sum_{i \cap r} t_i \cos \theta_i
 \ee 
 In which $i$ is the filament index and $\theta_i$ is the local angle between the filament and the radial direction. For a single simulation contracting with time, we find that the active stress indeed scales linearly with the density, as the density increases over time, Fig. \ref{illus_density}, inset. The linear stress-density relation also holds for an array of simulations with different initial densities, Fig. \ref{illus_density}. Interestingly, because $\alpha$ is constant in time, this also implies that viscosity scales linearly with actin density (Eq. \ref{pred_alpha}).

While we have confirmed the scaling of contractile stress with density, we do not know yet what sets the motor force $f_m$ in Eq. \ref{stress_fm}, nor the motor velocity $v_m$. Indeed, while motor force-velocity equations (Eq. \ref{motor_1}--\ref{motor_3}) should hold, we are lacking an additional constraint to set the motor velocity.

Qualitatively, we expect another constraint : if motor speed increases, the network should undergo more constractile stress, and thus motors should be under more tension, yielding a monotonically increasing relation between $f_m$ and $v_m$. The actual motor speed and force should be the intersection between this linear relation and the motor force-velocity relation, Eq. \ref{motor_1}. In the following, we give a scaling argument for this constraint.

To find a mean-field prediction for the force and velocity at which motors operate, let us consider the power per unit surface $\mathcal{P}$ produced by contractile units in the network, and the power $\mathcal{D}$ dissipated by the network viscosity . The former should scale as the density of force-producing units times force times velocity, the second as the viscous stress (scaling like $\eta \alpha$)  times the contraction rate $\alpha$, yielding :
\begin{eqnarray}
\mathcal{P} \propto \rho_u f_m v_m \mathcal{C} \label{power_p} \\
\mathcal{D} \propto \eta \alpha^2 \label{power_d}
\end{eqnarray}

Let us remember that $\alpha \propto \sigma^a / \eta$ and that $\sigma_a \propto \phi f_m \mathcal{C}$. Assuming that most of the contractile power is dissipated through viscosity, $\mathcal{P}\approx \mathcal{D}$, yields $v_m = f_m / \gamma_e$, in which $\gamma_e \propto \eta \rho_u / \phi^2 \mathcal{C}$ is an effective drag. Combining this with Eq. \ref{motor_1}, we find : 
\be
f_m = v_0 \left( \frac{1}{\gamma_e} +\frac{v_0}{f_s} \right)^{-1} \label{force_motor}
\ee

\begin{figure}[t]
	\includegraphics[width=7cm]{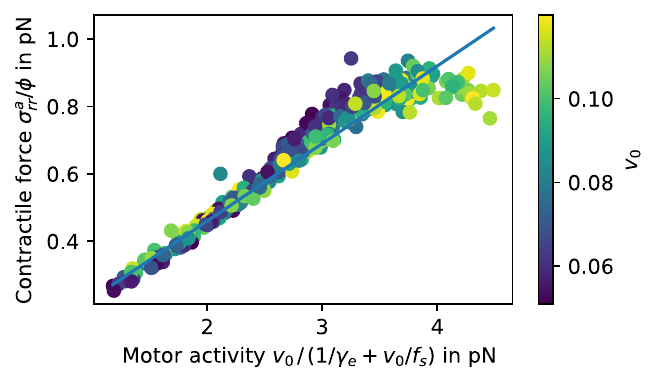} 	
	\caption{\label{illus_force} \small Contractile force $\sigma_{rr}^a / \phi$ for various simulations as a function of motor activity  $v_0 / ( 1/\gamma_e + v_0/f_s)$, where both $v_0$ and $f_s$ were varied systematically. Each point is one simulation, and color corresponds to motor unloaded speed $v_0$. $\gamma_e = 100 pN \mu m^{-1} s$ was fitted by aligning the data to a single line. Blue line is a linear guideline. }
\end{figure}

To test this result, we ran simulations where we systematically varied motor properties $f_m$ and $v_0$. We found that $\sigma_{rr}^a / \phi$  -- that should be proportional to $f_m$, see Eq. \ref{stress_fm} -- does scale as predicted by Eq. \ref{force_motor}, see Fig. \ref{illus_force}.

\begin{figure}[t]
	\includegraphics[width=7cm]{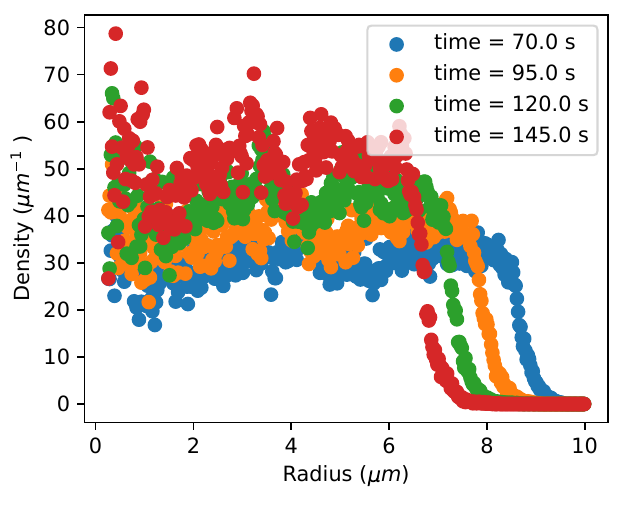} 	
	\includegraphics[width=8cm]{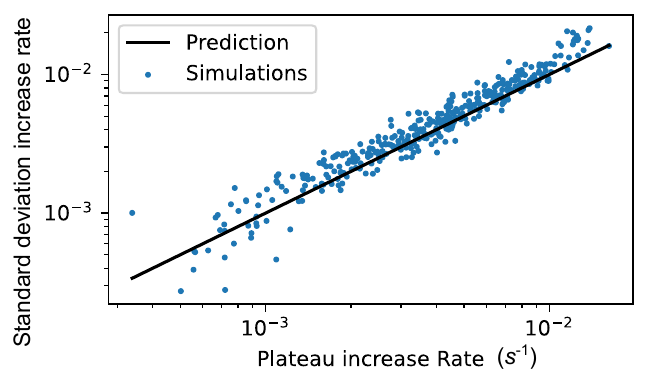} 	
	\caption{\label{illus_fluct} \small Top : representative example of actin densities at different times during a simulation. Bottom : rate of increase of actin intensity standard deviation $\partial_t \sqrt{< (\phi - \bar{\phi})^2 >_r}/ \bar{\phi}$ as a function of the rate of increase of the density plateau $\partial_t \bar{\phi} / \bar{\phi} $. Blue dots are simulations for different motor activities (see fig. \ref{illus_force}). Black line represents the prediction that both rates are equal. }
\end{figure}

\paragraph*{Stability of the contraction process}
The density of actin in the dense phase exhibits peaks and fluctuations, visible in the large standard deviation in Fig. \ref{illus_active_gel}. Individual simulations reveal that density profiles are heterogeneous, and heterogeneities seem to grow with time, Fig. \ref{illus_fluct}, top. We wondered if this apparent instability was intrinsic to the mechanics of contractile visco-elastic systems, in agreement with previous studies \cite{bois2011pattern, banerjee2011instabilities}.

We first performed a linear stability analysis on the system dynamics, Eqns. \ref{eq_stress},\ref{eq_mass}. Using the linearity of contractile stress and viscosity with respect to density, and assuming $\phi(r,t) = \bar{\phi}(t)  \left(1 + \epsilon \Phi(r,t) \right)$ and $v(r,t) = - \alpha r + \epsilon V(r,t)$, we find : 
\begin{eqnarray}
\epsilon \left[V''(r,t) + \frac{1}{r} V'(r,t) - \frac{1}{r^2}V(r,t) \right] = 0 + O(\epsilon^2)
\end{eqnarray}
Thus at first order, velocity is independent of density heterogeneities $\Phi(r,t) $. Heterogeneities are therefore transported with the matter flux. It is thus convenient to write transport in the Lagrangian frame of reference, describing the position of an element as $r \equiv r(R,t)$, with $R$ the initial position of the element. At first order, $v\propto r$ and contraction is homogeneous, yielding :
\begin{eqnarray}
r(R,t)= e^{-\alpha t} R , \qquad 
\phi(R,t)= \phi_0 (R) e^{ 2 \alpha t}
\end{eqnarray}
with $\phi_0$ the initial density profile. Thus, with a homogeneous contraction process, the entire density grows with the same rate $2 \alpha$ because of mass conservation. We therefore decided to go back to our simulations results, and compare the growth rate of heterogeneities (measured by the standard deviation of the density profile) to the growth rate of the plateau (measured by the mean density). We found that simulations did behave as predicted for actin gels, with density heterogeneities growing at the same rate as the mean, Fig. \ref{illus_fluct}, bottom. 
	
We showed that in simulations, both stress and viscosity are linear with actin density. We found that in that case, both in simulations and for the active model, fluctuations grow at the same speed as the mean density. However, we could also generically write active gel models with non-linear dependencies (for instance such as power laws :  $\sigma_{rr}^a = \zeta \phi^n$ and $\eta = \eta_0 \phi^m$). In this case, we could show that it is possible to get stable contraction (if viscosity dominates) or unstable contraction (if contractility dominates), see Supplementary Information. Actin systems sit at the stability transition $m=n=1$, and are thus marginally stable.


Active gel models are a popular model for cytoskeleton networks as they provide a clear explanation of possible phase transitions in active matter. They offer the simplicity of having a restricted number of macroscopic parameters rather than relying on large microscopic recipes. However, this aspect also often prevent them from quantitatively predicting experimental results. 

In this article, we showed that active gel models are a good macroscopic description of contractile actin-myosin networks, provided that contractility and viscosity depend on actin density. We showed that we can expect the contractile stress to be linear with actin density, as confirmed by our simulations. We also predicted how contractile stress depends non-linearly on motor stall force and unloaded speed, which we could validate. 

This is a first step towards a multi-scale understanding of actin mechanics. Once this module has been validated, it becomes possible to add other elements on top. For instance, we did not consider filament turnover here to single out contraction by motors. The balance between contraction and turnover could result in a transition from homogeneous to heterogeneous networks \cite{torres2010filament} ; this could be a key factor in blebbing, where the cell membrane detaches from the actin network and creates a protrusion \cite{kelkar2020mechanics}. Moreover, there could be a qualitative difference in large-scale behaviors of networks contracting because of motor activity versus those contracting due to actin disassembly \cite{bun2018disassembly}. It would now be possible to revisit this difference with a well established active gel model that adequately takes into account the microscopic origin of contractility.

In addition, we found that contractile actin was marginally stable, sitting at the boundary between the stable and the unstable regime. Thus, in cells, actin could be made to be finely tunable : adding additional ingredients could make viscosity dominate, and lead to homogeneous system, or make contractility dominate, and lead to strongly heterogeneous systems. Fine local tuning of a marginally stable system could therefore be a key to understanding the great diversity of actin structures observed within cells \cite{letort2015dynamic}.


\section{Acknowledgements}
The authors would like to thank Nicolas Minc, Martin Lenz, Jean-François Joanny for scientific discussions. We gratefully acknowledge Joel Marchand for the IT support and the IPOP-UP computing cluster. S.D. would like to thank the INSERM Aviesan grant “MMINOS” and Université Paris Cité’s Émergence program for financial support. D.W. received funding from European Research Council (ERC) grant ERC-SyG 101071793.  S. C. was supported by a SNSF project grant 200021-197068 to Guillaume Salbreux.

\section{Data accessibility} All code used in this work will be publicly released upon publication.

\bibliographystyle{unsrt}
\bibliography{bibtex_contractile}

\renewcommand\thefigure{S.\arabic{figure}} 
\setcounter{figure}{0} 

 \onecolumngrid

\section*{Supporting information}

\subsection*{Setup and equations from paper}

The governing equations in Eulerian coordinates read: 

\begin{equation}
	2 \del_r(\eta(\phi)\del_r v) + \frac{2 \eta(\phi)}{r}(\del_r v - \frac{v}{r}) +\del_r \sigma_{rr}^a = 0
\end{equation}
and
\begin{equation}
	\del_t \phi = -\frac{1}{r}\del_r (r v \phi)\label{eq:mass_conserv_eulerian}
\end{equation}

The boundary conditions should be $v(r = 0) = 0$ and $\sigma_{rr}^p+\sigma_{rr}^a\lvert_{\text{free bndry}} = 0$.
Further we assume a general power law for $\eta = \eta_0 \phi^m$ and $\sigma_{rr}^a = \zeta \phi^n$. 

\subsection*{Transformation to Lagrangian frame of reference}
The first step would be to express the force balance equation in terms of Lagrangian coordinates i.e. with respect to a reference state which will be the state of the gel at $t = 0$. 
The mapping is defined through 
$r \equiv r(R,t)$, $\phi \equiv \phi(R,t)$, together with a Jacobian $J(R,t) = \frac{r}{R}\del_R r$ which expresses the local volume (area) change. 
The density can be written as $\phi(R,t) = \phi_0/J(R,t)$ which automatically ensures mass conservation. 
$J<1$ means local compression, hence density increases and $J>1$ corresponds to local dilation which lowers the local density.
The local velocity in the gel is now defined as the time derivative of the mapping $v \equiv \del_t r(R,t)$. 
All derivatives transform according to the chain rule via $\del_r (.) = \frac{1}{\del_R r}\del_R (.)$\\

Applying these transformations allows to write the force balance equation in terms of Lagrangian coordinates as
\begin{equation}
	\del_R \open 2 \eta_0 \phi^m \frac{\del_R \del_t r}{\del_R r} + \zeta \phi^n \close + 2 \eta_0 \phi^m \open \frac{\del_R \del_t r}{r} - \frac{\del_R r \del_t r}{r^2}\close = 0\label{eq:fb_lagrange}
\end{equation}
The boundary conditions can be expressed as 
$\del_t r(R,t)\lvert_{R = 0} = 0$ and $\sigma_{rr}^p+\sigma_{rr}^a\lvert_{R = R_0} = 0$. 
For the stress free boundary I find 
\begin{equation}
	\open\frac{\del_R \del_t r}{\del_R r}+ \frac{\zeta}{2\eta_0}\phi^{n-m}\close\Big\lvert_{R = R_0}=0\;.\label{eq:stress_free_boundary_general}
\end{equation}

\subsection*{Homogenous contraction}

The case of homogenous contraction would be written as $r_0(R,t) \equiv a(t)R$. It leads to a contraction which keeps the density homogenous.
This can be derived as follows: 
The density $\phi(R,t) = \phi_0 \frac{R}{r \del_R r}$ remains homogenous if $\phi(R,t) = \phi_0 \psi(t)$. 
This allows to solve 
\begin{equation}
	\frac{R \diff R}{\psi(t)} = r \diff r\;,
\end{equation} 
which has the general solution 
\begin{equation}
	r(R,t) = \sqrt{\frac{R^2}{\psi(t)}+2 C(t)}\;,
\end{equation}
with integration constant $C(t)$.
We demand that $r(0,t) = 0$ such that $C(t) = 0$ and $r(R_0,0) = R_0$ such that $\psi(0) = 1$. with $a(t)\equiv \psi(t)^{1/2}$ we find 
\begin{equation}
	r(R,t) = a(t)R\;,
\end{equation}
with $a(0) =1$.

It can be easily verified that $r_0(R,t)$ solves the PDE. 
The stress free boundary condition for this case reads 
\begin{equation}
	\open \frac{\del_t a}{a} + \frac{\zeta}{2 \eta_0}(\frac{\phi_0}{a^2})^{n-m}\close\Big\lvert_{R = R_0} = 0
\end{equation}
from which one can obtain the time evolution by solving the ODE for $a(t)$. 
In the special case $n = m$ we find
 \begin{equation}
	a(t) = e^{-\frac{\zeta}{2\eta_0} t}\;.
\end{equation}

\section*{Small perturbations around homogenous state}
We now perturb around this homogenous state by assuming
\begin{equation}
	r(R,t) \approx r_0(R,t) + \delta r(R,t) = a(t) R + \epsilon u(R,t)
\end{equation}
with $\epsilon \ll 1$.

First we will expand all terms in \ref{eq:fb_lagrange} up to order first order in $\epsilon$:

\begin{equation}
		\phi(R,t) = \frac{\phi_0 R}{r\del_R r} \approx \frac{\phi_0}{a^2}\open 1 - \epsilon \frac{R \del_R u + u}{a R}\close\;,\label{eq:density_of_u}
\end{equation}
from which we find 
\begin{equation}
		\phi^m \approx \open\frac{\phi_0}{a^2}\close^m \open 1 - \epsilon m \frac{R \del_R u + u}{a R}\close\;,
\end{equation}
\begin{equation}
		\frac{\del_R \del_t r}{\del_R r} \approx \frac{\del_t a}{a}+\epsilon \open \frac{\del_R \del_t u}{a}-\frac{\del_t a}{a^2}\del_R u\close\;,
	\end{equation}

\begin{equation}
		\frac{\del_R \del_t r}{r} \approx \frac{\del_t a}{a R}+\epsilon \open \frac{\del_R \del_t u}{a R}-\frac{u \del_t a}{(a R)^2}\close\;,
\end{equation}
\begin{equation}
		\frac{\del_R r \del_t r}{r^2} \approx \frac{\del_t a}{a R} + \epsilon \open \frac{R \del_t a \del_R u + a \del_t u - 2 u \del_t a}{(a R)^2}\close
\end{equation}

The force balance equation for a general $\eta(\phi)$ and $\sigma_a(\phi)$ can be written as 
\begin{equation}
	2 \del_R \phi \open \del_\phi \eta(\phi) \frac{\del_R \del_t r}{\del_R r} + \del_\phi \sigma_a(\phi)\close + 2 \eta(\phi)\open \del_R \open \frac{\del_R \del_t r}{\del_R r}\close + \frac{\del_R \del_t r}{r} - \frac{\del_R r \del_t r}{r^2}\close = 0\;.
\end{equation}
Plugging in all the linearized terms and keeping only terms of $\mathcal{O}(\epsilon)$ we find that the force balance equation reduces to a time-evolution equation of the operator
\begin{equation}
	\mathcal{L}[u(R,t)] \equiv -u + R \del_R u + R^2 \del_R^2 u
\end{equation}
which can be expressed as 
\begin{equation}
	\del_t \mathcal{L}[u(R,t)] = \open (m+1)\frac{\del_t a}{a}+n \frac{\zeta}{\eta_0}\open\frac{\phi_0}{a^2}\close^{n-m}\close \mathcal{L}[u(R,t)] = \lambda(t) \mathcal{L}[u(R,t)]\;,\label{eq:time_evolution_operator}
\end{equation}
and is solved by 
\begin{equation}
	\mathcal{L}[u(R,t)] = \mathcal{L}[u(R,0)]e^{\int_0^t \lambda(s)\diff s}\;.\label{eq:time_evolution}
\end{equation}
Hence, the time evolution of the operator $\mathcal{L}$ is fully determined by $\lambda(t)$.
Together with the boundary conditions from the zeroth order (homogeneous constraction) we identify 
\begin{equation}
	\lambda(t) = \open n - \frac{m+1}{2}\close\frac{\zeta}{\eta_0}\open \frac{\phi_0}{a^2}\close^{n-m}\;.
\end{equation}

To gain insight into the time evolution of $u(R,t)$ we first note that the kernel of $\mathcal{L}[w(R,t)] = 0$ is given by
\begin{equation}
	w(R,t) = c(t)R + d(t) R^{-1}\;.
\end{equation} 
$d(t)$ must vanish to ensure regularity at $R = 0$. The remaining (kernel) solution is not a ``physical'' perturbation because it leads to $r(R,t) = (a(t) + c(t))R$ which necessarily demands $c(t) = 0$ to ensure the stress free boundary already satisfied by $a(t)$. 

Because $\mathcal{L}$ has no explicit time dependence, we can use $\del_t \mathcal{L}[u] = \mathcal{L}[\del_t u]$ and write 
\begin{equation}
	\mathcal{L}[u - \lambda(t)u] = 0\;.
\end{equation}
Restricting ourselves to solutions $u(R,t)$ which do not lie in the kernel of $\mathcal{L}$ we obtain the unique evolution law
\begin{equation}
	\del_t u = \lambda(t) u \implies u(R,t) = u(R,0) e^{\int_0^t \lambda(s)\diff s}\;.
\end{equation}
Hence, in Lagrangian coordinates the spatial shape of each physical perturbation is fixed, and only its amplitude changed in time. The growth is determined by the sign of $\lambda(t)$ with the criterion that perturbations grow if the active stress is sufficiently strong i.e. for $n > (m+1)/2$.

\begin{figure*}
	\includegraphics[scale=1.5]{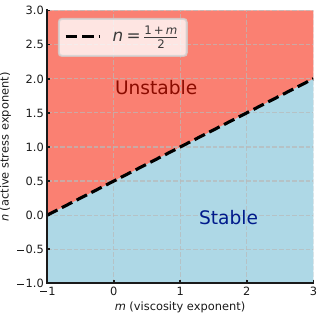}
    \caption{Stability diagram for deformation perturbations $u(R,t)$. Depending on the exponents in the viscosity $\eta(\phi) = \eta_0 \phi^m$ and active stress $\sigma_a(\phi) = \zeta \phi^n$ perturbations $u(R,t)$ grow or shrink in time. }
	\label{fig:phase}
\end{figure*}

\end{document}